\documentclass[runningheads]{llncs}
\usepackage{graphicx}
\usepackage{longtable}
\usepackage{booktabs}
\begin{document}
\title{mobilityDCAT-AP: a Metadata Specification for Enhanced Cross-border Mobility Data Sharing}
\titlerunning{mobilityDCAT-AP: Metadata for Enhanced Mobility Data Sharing}

%
\author{Mario Scrocca\inst{1}\orcidID{0000-0002-8235-7331} \and
Lina Molinas Comet\inst{2}\orcidID{0000-0001-5446-6947} \and
Benjamin Witsch\inst{3}\orcidID{0009-0001-4110-3802} \and
Daham Mohammed Mustafa\inst{2}\orcidID{0000-0003-1867-4428} \and
Christoph Lange\inst{2,4}\orcidID{0000-0001-9879-3827} \and
Marco Comerio\inst{1}\orcidID{0000-0003-3494-9516} \and
Peter Lubrich\inst{5}\orcidID{0000-0002-0023-1234}}
\authorrunning{M. Scrocca et al.}
%
\institute{Cefriel -- Politecnico di Milano, Milan, Italy
\email{\{mario.scrocca,marco.comerio\}@cefriel.com}\\ \and
Fraunhofer Institute for Applied Information Technology FIT, \\ Sankt Augustin, Germany\\
\email{\{lina.teresa.molinas.comet,christoph.lange-bever,\\daham.mohammed.mustafa\}@fit.fraunhofer.de}\\ \and
AustriaTech, Vienna, Austria \\
\email{benjamin.witsch@austriatech.at} \and
RWTH Aachen University, Germany \and
Federal Highway Research Institute (BASt), Bergisch Gladbach, Germany\\
\email{lubrich@bast.de}}
\maketitle              
\begin{abstract}
Integrated and efficient mobility requires data sharing among the involved stakeholders. In this direction, regulators and transport authorities have been defining policies to foster mobility data's digitalisation and online publication. However, creating several heterogeneous data portals for mobility data resulted in a fragmented ecosystem that challenges data accessibility. In this context, metadata is a key enabler to foster the findability and reusability of relevant datasets, but their interoperability across different data portals should be ensured. Moreover, each domain presents specificities on the relevant information that should be encoded through metadata. To solve these issues within the mobility domain, we present mobilityDCAT-AP, a reference metadata specification for mobility data portals defined by putting together domain experts and the Semantic Web community. We report on the work done to develop the metadata model behind mobilityDCAT-AP and the best practices followed in its implementation and publication. Finally, we describe the available educational resources and the activities performed to ensure broader adoption of mobilityDCAT-AP across mobility data portals. We present success stories from early adopters and discuss the challenges they encountered in implementing a metadata specification based on Semantic Web technologies.  \\
\\
\textbf{Resource type}: Metadata Specification \\ 
\textbf{License}: Creative Commons Attribution 4.0 License \\
\textbf{PID}: https://w3id.org/mobilitydcat-ap \\
\textbf{Repository}: https://github.com/mobilityDCAT-AP

\end{abstract}
\section{Introduction}

Digitalisation in the mobility domain enabled the collection of heterogeneous data assets from various stakeholders that could leverage them to develop data-driven applications, ranging from dashboards for data visualizations to AI-enabled algorithms for prediction and prescription~\cite{data_mobility}. Nevertheless, developing effective Intelligent Transportation Systems (ITS) and enabling Mobility-as-a-Service (MaaS) requires the collaboration of different mobility stakeholders and mandates for data-sharing solutions. In this direction, Web portals for mobility data have been developed worldwide in recent years promoting open data best practices~\cite{yadav2017role}. These portals often have specific spatial or thematic coverage, e.g., exposing data for one specific region only or containing only data about public transportation modes. As a result, data are scattered across several data portals. To address this issue, the European Commission required each Member State to establish and populate a National Access Point (NAP)~\cite{commission_2017} to facilitate access and reuse of mobility data.

Metadata is a crucial building block for the discoverability of datasets within NAPs and other mobility data portals. However, a common metadata specification is needed to enable interoperability and metadata exchanges across different portals~\cite{lubrich2021harmonised}.
The objective is to have homogeneous data descriptions that ease searching, discovering, and accessing the proper data resources through the relevant data portal. The problem to be addressed is twice. On the one hand, domain expertise is needed to identify the relevant information to be represented within metadata. On the other hand, a technological solution based on Semantic Web technologies should be adopted to enable semantic interoperability with widely adopted vocabularies based on DCAT~\cite{dcat2}.

This paper introduces a formal metadata specification for mobility data portals called mobilityDCAT-AP\footnote{The specification has been previously referred as napDCAT-AP~\cite{scrocca_towards_2022}, then changed to mobilityDCAT-AP to clarify its applicability to any mobility data portal.}, recently elaborated by a stakeholder and expert group in the EU-funded project NAPCORE\footnote{\url{https://napcore.eu/}} (National Access Point Coordination Organisation for Europe). 

mobilityDCAT-AP enables harmonised, platform-independent metadata descriptions that are both human-readable and machine-actionable formats. The latter ensures seamless integration of mobility-related platforms with third-party systems by enabling the automated import and export of metadata via, for example, an API. mobilityDCAT-AP defines precise and unambiguous fields to describe metadata relevant to mobility data sources, e.g. for representing the data topic, the data provider or the geographical coverage. mobilityDCAT-AP follows best practices for defining metadata specifications using Semantic Web technologies and has been designed to be interoperable with existing specifications. mobilityDCAT-AP is recommended for metadata management in National Access Points (NAPs) in Europe. Nevertheless, any other mobility data portals can leverage it to harmonise their data descriptions and ease the metadata exchange in the overall mobility data ecosystem. 

The remainder of this paper is structured as follows. Section~\ref{sec:preliminaries} discusses preliminaries on data-sharing in the mobility domain and related works.
Section~\ref{sec:methodology} discusses the methodology followed for designing, implementing and publishing mobilityDCAT-AP. This section describes the artefacts generated and released for mobilityDCAT-AP, but also provides guidelines for defining similar metadata specifications in other domains. Section~\ref{sec:maintenance} comments on maintenance and sustainability plans. Sections~\ref{sec:adoption} and \ref{sec:challenges} discuss the current status of the adoption of mobilityDCAT-AP, challenges and lessons learnt. Finally, Section~\ref{sec:conclusion} draws conclusions and presents future work.

\section{Preliminaries and Related Works}
\label{sec:preliminaries}

This section presents the current landscape of regulations and initiatives for data-sharing in the mobility domain, the metadata models for mobility data portals and NAPs, and the existing metadata specifications in other domains.

\subsection{Towards a Federation of Mobility Data Portals}

The European Commission, through the EU Delegated Regulation 2017/1926~\cite{commission_2017} and other regulations~\cite{Commission2013Delegated,CommissionRealTime,CommissionSSTP} supplementing Directive 2010/40/EU~\cite{Parliament2010Directive} 
with regard to the provision of EU-wide multimodal travel information services, has required transport authorities in each EU country to set up a National Access Point (NAP) to facilitate the sharing and reuse of transport and mobility data. 
A NAP represents a single national access point for data that can be used to develop new applications for innovative, efficient, and sustainable mobility. Data made available through NAPs are provided by transport authorities and operators and cover a wide range of information related to mobility services, traffic, and road safety.

The NAPCORE project aims to establish a long-lasting organisation for coordinating all NAPs in Europe on an organizational and technical level. The goals of NAPCORE include establishing unified recommendations concerning data exchange technologies, standards, formats, and processes used by the NAPs. It also aims to enhance interoperability for improved discoverability and accessibility of common data, thereby facilitating pan-European services. Additionally, NAPCORE seeks to formulate shared strategies that address existing and upcoming developments.
One of the problems addressed by the NAPCORE project is the appearance of different metadata vocabularies in European NAPs and the need for interoperability across NAPs. The problem of heterogeneous metadata descriptions is also reflected in inconsistent web services and interfaces offering limited search capabilities for users and ITS service providers. For this reason, the NAPCORE project worked to define mobilityDCAT-AP as the recommended metadata specification for NAPs across Europe.

Moreover, to facilitate the trusted exchanges of business-sensitive mobility data, the European Commission is promoting the development of common European data spaces in strategic economic sectors, including transport and mobility~\cite{commission_data}. 
The EU Communication on creating a common European mobility data space (EMDS)~\cite{commission_emds} highlights that data discoverability and accessibility must be facilitated by defining common metadata from existing and emerging mobility and transport data domains.

The Shift2Rail IP4 SPRINT project implemented an automated solution for ingesting and harmonising metadata from selected NAPs to enable cross-border multimodal journey planning~\cite{carenini_enabling_2021}. As a baseline, a conceptual mapping between different NAP metadata schemes was investigated. The adoption of mobilityDCAT-AP could streamline this scenario by avoiding the need for metadata conversion. The same metadata specification could enable for a more easily federation of mobility data portals with cross-portal search functionality.

\subsection{Metadata Specifications for Data Portals enabled by Semantic Web Technologies}

A metadata specification is a structured framework or standard that defines the information needed to describe a specific resource. When metadata is not provided, e.g., a data source is published without information, it is difficult to know important information like the date of publication or the license to reuse the data. Additionally, if metadata is not exposed in a machine-actionable (e.g., metadata only provided within a textual description on a website), they can not be processed automatically hindering data sharing and integration with other platforms.

In addition, even when metadata from different data portals can be exported, several challenges arise when considering their harmonization. First, there is the adoption of different terminology for the same information, for example the terms “supplier” and “publisher” may be used to represent the same metadata field. Furthermore, using varying sets of metadata fields presents another challenge, as certain specific information may not be available in all data portals. Lastly, the application of different values for metadata fields, e.g., custom string values used as data categories, makes it difficult to group similar data sources together. 

To address these challenges, Semantic Web technologies provide a practical solution by encoding semantics in an interoperable, machine-actionable format. To tackle the challenge of differing terminology, reference vocabularies defined in RDF and published online can ensure uniformity of terminology and semantics for metadata fields. For the challenge related to inconsistent metadata sets, application profiles can be established to enhance interoperability by outlining constraints and expected combinations regarding using one or more vocabularies as metadata fields. To resolve the issue of varied values for metadata fields, controlled vocabularies, e.g., SKOS~\cite{miles2009skos} taxonomies, can ensure that the same entity represents the same value for a particular metadata field.

The Data Catalog Vocabulary (DCAT)~\cite{dcat2} is a widely adopted RDF vocabulary designed to describe data catalogues using a standardized set of classes and properties to model data sources. 
The DCAT Application Profile (DCAT-AP)~\cite{dcatap}
specifies, considering DCAT as the base vocabulary, a profile for data portals in Europe to favour the aggregation, exchange, search and automated processing of metadata.  DCAT-AP defines cardinalities and obligations (mandatory, recommended, and optional) for DCAT elements. Moreover, it provides recommendations for controlled vocabularies\footnote{https://op.europa.eu/en/web/eu-vocabularies/controlled-vocabularies} and additional properties that can be used for metadata. The implementation of mobilityDCAT-AP is currently aligned with DCAT version 2 and DCAT-AP version 2.0.1.

DCAT-AP scope is cross-border and cross-domain, and thus further specified in different, domain-specific extensions like GeoDCAT-AP~\cite{geodcat-ap} for spatial data and StatDCAT-AP~\cite{statdcat-ap} for statistical datasets. Examples of how these extensions adapt DCAT-AP to their specific requirements are: (i) extension of controlled vocabularies to be used as recommended values for metadata properties, (ii) definition of additional metadata properties (e.g., to describe the quality process, legal details to access a resource), and (iii) changes in cardinalities and obligations for metadata properties.

DCAT-AP and its extensions enhance the interoperability and discoverability of data portals. Data portals that present metadata aligned with DCAT-AP allow applications to seamlessly access metadata from various sources. Additionally, this consolidated metadata can be utilized to support a comprehensive data portal, facilitating the findability and discoverability of data. One example for such consolidation is the European data portal\footnote{\url{https://data.europa.eu/}}), harvesting and exposing numerous metadata sets from national and regional open data portals in Europe. mobilityDCAT-AP defines a domain-specific extension of DCAT-AP, thus providing compatibility with data portals that support DCAT-AP metadata.


The SEMIC (Semantic Interoperability Community) established by the European Commission, targeting the public sector and beyond\footnote{\url{https://interoperable-europe.ec.europa.eu/collection/semic-support-centre}}) is responsible, among others, for the DCAT-AP specification. Furthermore, they provide best practices and support for the definition of DCAT-AP extensions. We leveraged the resources and support from SEMIC experts as reported in Section~\ref{sec:methodology}. 


\subsection{Metadata Specifications for Mobility Data}

The Coordinated Metadata Catalogue (CMC) of the EU EIP project (European ITS Platform) was the first common blue-print for metadata structures for NAPs in Europe~\cite{cmc}. Thus, it can be considered as a precursor of the present work. It defines a common minimum set of metadata, including descriptions, formatting constraints and obligation levels. However, these definitions are only available in a proprietary, human-readable format. The CMC does not provide guidelines on how to implement the metadata specification, thus neither promoting metadata interoperability nor semantic technologies. 

TransportDCAT-AP~\cite{transportdcatap} is an extension of DCAT-AP focusing on the public transport domain. 
The profile takes into account the geospatial nature of public transport data by defining the set of metadata related to this type of information as mandatory. In addition, a specific set of admissible keywords is defined to standardise the range of metadata properties (i.e., the values that can be associated with a property) and to enable domain-related queries, e.g., considering specific transportation modes.
Despite similar objectives to mobilityDCAT-AP, the extension was developed as part of a specific project and is currently not actively maintained or adopted by mobility data portals.

The INSPIRE implementing rules provide guidelines for metadata\footnote{\url{https://github.com/INSPIRE-MIF/technical-guidelines/tree/main/metadata}} in the context of spatial information, adopting the ISO/TS 19139-based XML format. The INSPIRE metadata scheme is in turn represented via another DCAT-AP extension\footnote{\url{https://semiceu.github.io/GeoDCAT-AP/releases/}}). As the spatial aspect is highly relevant in the mobility domain, we widely reused the work done by GeoDCAT-AP for defining mobilityDCAT-AP. This provides support and mappings for INSPIRE metadata.

\section{Methodology for a Metadata Specification Applied to the Mobility Domain}\label{sec:methodology}

Relying on existing literature and other metadata specifications, we defined a roadmap comprising five main steps for the design, implementation and publication of a metadata specification: (1) gathering requirements from experts and domain stakeholders, (2) defining a conceptual model, (3) implementing the conceptual model by adopting the Semantic Web stack, (4) documenting the specification and providing guidelines for its use, and (5) hosting and publishing all outputs.
To support the definition of a roadmap the following inputs were considered (i) documents from the literature describing DCAT-AP and its extensions, (ii) artefacts published online for DCAT-AP and its extensions, and (iii) interviews with DCAT-AP and SEMIC experts. Details of the roadmap and the requirements analysis are presented in~\cite{scrocca_towards_2022}. In this section, we summarise the results for the requirements specification and we discuss the complete application of the roadmap for mobilityDCAT-AP.

\subsection{Requirements Specification}

The initial phase of the roadmap focused on defining the requirements for the DCAT-AP extension. It was essential to identify a series of use cases and functionalities that the extension needs to support, engaging relevant stakeholders in this process to gather a comprehensive set of requirements. This phase also took into account current projects and initiatives related to metadata that are in the scope of the ontology (e.g., same domain) to ensure their principles are integrated into the design and implementation stages. This is also important to facilitate the adoption of the extension. Last but not least, existing DCAT-AP extensions were carefully reviewed since they may already cover relevant aspects for the development of the new extension.

For the requirements' definition, we executed a comprehensive literature analysis and collected contributions from relevant stakeholders, including NAPCORE partners, NAP developers, metadata providers, data portal users, and Semantic Web experts. The references examined encompass a range of sources, such as previous research, European legislation, metadata standards, and specific DCAT-AP materials. These resources are categorized into six main groups, notably open data literature, EU directives related to public sector information, metadata standards, and project results. Domain experts involvement followed a more practical collection of inputs around a basic question: \emph{what is essential information about a mobility data offering on a data portal?}.


As a result of the literature review and stakeholder input, NAPCORE identified forty requirements for mobilityDCAT-AP, which fall into four categories: \emph{General} (8), \emph{Existing Vocabularies} (5), \emph{Content} (22), \emph{Implementation} (5). The \emph{General} requirements outline high-level specifications, while \emph{Existing Vocabularies} pertain to existing initiatives to be considered. 
A detailed report of the literature reviewed, stakeholder insights, and the specific requirements defined is documented in the NAPCORE report \cite{mobilitydcataprequirements}.

\subsection{Conceptual Model}

The second step of the roadmap concerns defining the conceptual model and should follow the best practices provided by SEMIC~\cite{processdcat}.
A conceptual model is a simplified representation of a system, process, or idea that aids understanding by focusing on its essential components and relationships. This step corresponds to usual procedures for building data models, i.e., looking at business requirements as formulated by experts in the domain concerned and translating these requirements into a technology-independent data structure~\cite{datamodelingessentials}.
Relevant actions to be performed are: (i) analyse the requirements of the model to be developed, (ii) reach a consensus by developing semantic agreements among different stakeholders, (iii) define and implement a structured and clear model, and (iv) publish the domain model for public revision.

For the definition of mobilityDCAT-AP, we developed a conceptual model involving data platform operators, mobility stakeholders and experts from the Semantic Web community. 
Being related to the development of a metadata extension for DCAT-AP, we strongly relied on its specification to draft the initial version of the conceptual model. Additionally, a strong emphasis was given to previous harmonisation efforts such as the Coordinated Metadata Catalogue~\cite{cmc}. On these bases, the conceptual model was refined to address our requirements.

The essence of the conceptual model for a metadata specification consists of two main parts: the entities with properties and the related controlled vocabularies. The defined entities are based on crucial design decisions such as “Is there a use case for this entity?” or “Is an entity for basic or advanced information?”. This second aspect is of importance to then assign cardinality to each entity, i.e., optional, recommended or mandatory. After the creation of the initial entity set, the controlled vocabularies were developed based on active and planned regulations, experience and existing vocabularies from related domains such as tourism, logistics or geography. 

After defining the conceptual model, we generated two main artefacts: a class diagram to provide an overview and an Excel spreadsheet with a more detailed description of each entity and proposed values for each controlled vocabulary.
These two artefacts were used for a multi-stage review process with NAP operators, platform developers and other stakeholders. Further refinements of the conceptual model were necessary during the roadmap's implementation step to align it with existing extensions or vocabulary adopted for reuse.

\subsection{Implementation}


The third step in the roadmap focuses on implementing the established conceptual model in RDF. This phase involves identifying and/or defining the RDF classes and properties necessary to represent the domain model. Additionally, it includes the identification or implementation of controlled vocabularies defining the acceptable values for certain properties. Lastly, this stage addresses the generation of artefacts for the encoding of the specification in RDF.

\begin{figure}
     \centering
     \includegraphics[width=\linewidth]{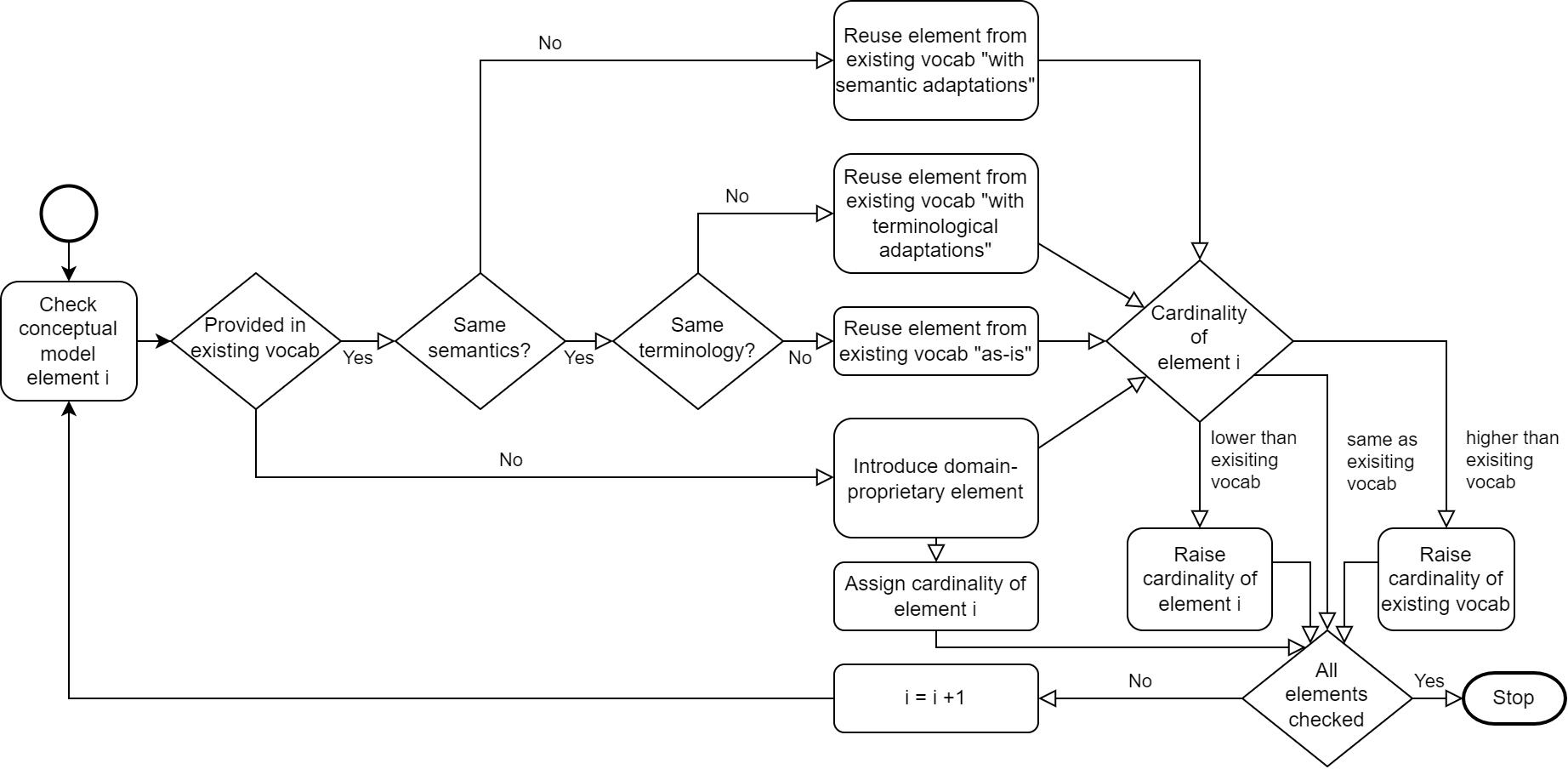}
     \caption{Flowchart for defining the RDF implementation of mobilityDCAT-AP}
     \label{fig:flowchart}
 \end{figure}

The implementation of mobilityDCAT-AP followed a structured process to transform the conceptual model into an RDF data model, as shown in Fig.~\ref{fig:flowchart}. This involved mapping to existing RDF vocabularies and defining additional terms for mobility-specific concepts not covered by existing vocabularies.

The implementation began with analyzing two categories of RDF vocabularies: (1) vocabularies already included in DCAT-AP and relevant extensions (e.g., GeoDCAT-AP for spatial information), such as DCMI Metadata Terms\footnote{\url{https://www.dublincore.org/specifications/dublin-core/dcmi-terms/}}, SKOS~\cite{miles2009skos}, and the Data Quality Vocabulary\footnote{\url{https://www.w3.org/TR/vocab-dqv/}}, and (2) other existing vocabularies that could address mobility-specific features, such as the ISA Programme Location Core Vocabulary\footnote{\url{https://github.com/SEMICeu/Core-Location-Vocabulary}}. 

The first step was to identify if one domain element (i.e., one relevant element of mobility metadata) was already covered by an existing vocabulary. If this was not the case, a domain-proprietary element was introduced, i.e., an RDF term in the namespace bound to the \texttt{mobilitydcatap:} prefix.
If an existing vocabulary did cover this domain element, multiple checks had to be made. First, it was checked if there is a match in terms of semantics and terminology, when comparing the domain element and the existing vocabulary. A terminological adaption would include a refining of the usage notes. A semantic adoption would imply creating a sub-class or a sub-property to the existing vocabulary, with a semantic description tailored to the mobility domain. If both semantics and terminology match, the existing vocabulary could be reused “as-is”. For example, mobilityDCAT-AP introduces a class \textit{mobilitydcatap:MobilityDataStandard} as a subclass of the existing class \textit{dct:Standard} due to semantic differences in both concepts. This procedure is in line with the SEMIC guidelines for generating domain-specific application profiles of existing RDF vocabularies~\cite{styleguide}. 
Second, the cardinality proposed in the conceptual model and the one in DCAT-AP was compared. As mobilityDCAT-AP is a formal extension of DCAT-AP, the DCAT-AP extensions rules~\cite{extensionrules} are to be strictly followed, implying that a cardinality can only be increased, but not lowered by an extension
An example in mobilityDCAT-AP was the \textit{dcat:accrualPeriodicity} property for the \textit{dcat:Dataset}, which was raised to “mandatory” according to the stakeholders' requirements.



As a result of this work, we implemented an OWL serialisation of new classes and properties not covered by DCAT-AP but deemed as needed to describe mobility data sources in mobilityDCAT-AP. 
We provided the implementation of the introduced classes and properties in the JSON-LD, Turtle, and RDF/XML serializations of RDF.

The implementation of mobilityDCAT-AP was complemented by the work related to controlled vocabularies. Similarly to the strategy described above, we aimed to reuse existing vocabularies whenever possible, following the recommendations defined in DCAT-AP and GeoDCAT-AP. Furthermore, we analysed controlled vocabularies maintained by the EU Publication Office \footnote{\url{https://op.europa.eu/en/web/eu-vocabularies/controlled-vocabularies}}. We identified and implemented eleven new controlled vocabularies to be defined for mobilityDCAT-AP: Application Layer Protocol, Communication Method, Conditions for Access and Usage, Mobility Theme, Mobility Data Standard, Georeferencing Method, Grammar, Network Coverage, Intended Information Service, Transport Mode, Update Frequency. Additional details on the recommendation for controlled vocabularies identified can be found within the mobilityDCAT-AP specification\footnote{\url{https://mobilitydcat-ap.github.io/mobilityDCAT-AP/releases/index.html\#controlled-vocabularies-to-be-used}}. The implementation of controlled vocabularies was aligned with other taxonomies referenced by DCAT-AP and follows the SKOS specification. We adopted the tool \texttt{xls2rdf}\footnote{\url{https://xls2rdf.sparna.fr/rest/doc.html}} for their maintenance and serialisation of controlled vocabularies directly from an Excel description. Indeed, a wide involvement of domain stakeholders was needed to define controlled vocabularies and the Excel spreadsheets greatly facilitated the collection of feedback from them.

\begin{figure}
    \centering
    \includegraphics[width=0.9\linewidth]{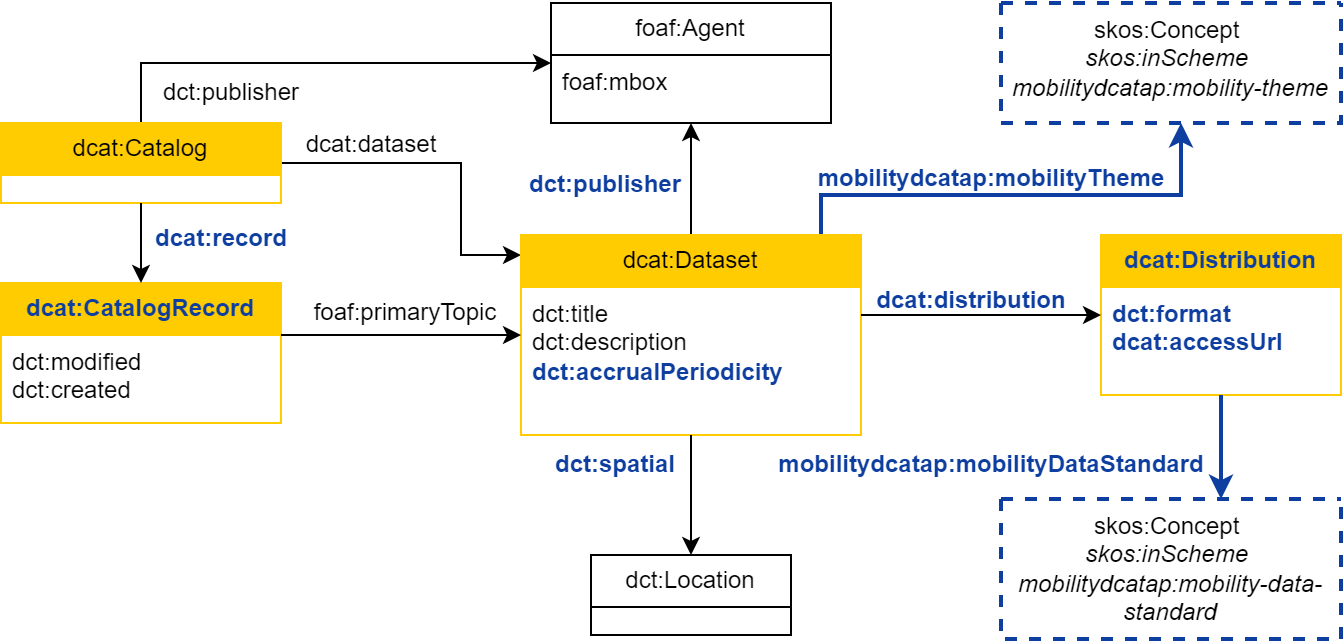}
    \caption{Minimum profile for mobilityDCAT-AP metadata}
    \label{fig:minimum-profile}
\end{figure}

Figure~\ref{fig:minimum-profile} shows the minimum profile for mobilityDCAT-AP considering only mandatory fields. The classes and properties in blue represent the modifications with respect to DCAT-AP. Dotted boxes refer to the controlled vocabularies to be used as values for the properties referencing them.

Before finalising the implementation of mobilityDCAT-AP, we involved experts from SEMIC working on DCAT-AP. The provided feedback helped us improve the alignment among the specifications and clarify different design decisions. Furthermore, we were invited to join SEMIC activities to ensure a better alignment also in the evolution of mobilityDCAT-AP.
 
\subsection{Documentation and guidelines}

The fourth step of the roadmap focused on outlining the documentation and supplementary materials that support the publication of the DCAT-AP extension. 
These supplementary materials include guidelines for implementing and using the specification, canonical mappings related to other pertinent metadata standards, and resources or tools for validating metadata that utilizes the DCAT-AP extension.

\paragraph{Documentation} The documentation of a metadata application profile slightly differs from that of an ontology since it may introduce new concepts but emphasizes practical implementation aspects considering existing vocabularies (e.g., entities to be used and cardinality constraints). Furthermore, it provides recommendations for usage, such as those associated with controlled vocabularies for specific properties. These specifications are provided through detailed textual documentation that complements the machine-readable serialization. Having evaluated WIDOCO, the Wizard for Documenting Ontologies~\cite{garijo2017widoco}, and the LinkML Linked Data Modeling Language~\cite{moxon_linked_2021}, which includes tools to generate documentation, we finally settled on using the ReSpec tool for online recommendations and specifications. We found the structure of the documentation generated by WIDOCO too specific to ontologies and not easily adaptable to the one required by mobilityDCAT-AP, whereas the main challenge with LinkML would have been the usage of an independent modelling language, requiring the definition of an additional intermediate representation of the new classes and properties, but also of all the vocabularies referenced by the specification.
ReSpec met our requirements; our decision for it was additionally motivated by the related GeoDCAT-AP already using it and SEMIC recommending ReSpec as a best practice. A custom style was defined for mobilityDCAT-AP specifications to include specific visual elements and accommodate changes to the structure of the web page\footnote{\url{https://github.com/mobilityDCAT-AP/respec-style}}.
Then, the ReSpec tool was used to generate documentation for working drafts and release versions of mobilityDCAT-AP.
For controlled vocabularies, an automatic workflow\footnote{\url{https://github.com/cefriel/controlled-vocabulary-template}} was established to support the generation of documentation directly from their Excel descriptions. WIDOCO~\cite{garijo2017widoco} is executed on the RDF serialization generated by \texttt{xls2rdf} to obtain RDF serializations of the Excel content and human-readable documentation.

\paragraph{Validation} For the automatic validation of metadata adopting mobilityDCAT-AP, we encoded recommendations and constraints defined by the specification (e.g., cardinality constraints) as SHACL shapes\footnote{\url{https://www.w3.org/TR/shacl/}} serialized in Turtle. These SHACL shapes import the ones from DCAT-AP to avoid repetitions and to ensure backward compatibility of mobilityDCAT-AP metadata with DCAT-AP. Any tool able to validate SHACL shapes can be used to validate mobilityDCAT-AP metadata against the provided shapes.

\paragraph{Resources for adopters} The release of mobilityDCAT-AP was accompanied by several resources to help stakeholders in the adoption and implementation of the metadata specification. Indeed, many stakeholders in the mobility domain are not familiar with Semantic Web technologies and should be helped in understanding the core technologies, the benefits of adopting a metadata specification like mobilityDCAT-AP and how to implement it. The main resource in this regard is the mobilityDCAT-AP Wiki\footnote{\url{https://github.com/mobilityDCAT-AP/mobilityDCAT-AP/wiki}}. The content, published in the form of Frequently Asked Questions (FAQ), is constantly updated to reflect feedback and questions we receive from adopters. The Wiki includes implementation guidelines and additional explanations on using specific properties. Furthermore, it provides: (i) guidelines for the validation of metadata, explaining the main concepts of SHACL and how to validate RDF using a set of shapes, (ii) details on the governance and maintenance of the specification, (iii) pointers to relevant artefacts to help adopters such as examples of metadata descriptions using mobility DCAT-AP and mappings from the CMC metadata standard to mobilityDCAT-AP.
In addition to this content, NAPCORE organise webinars and events to disseminate the specification and provide support to implementers. An additional tool made available is the \emph{mobilityDCAT-AP Generator}\footnote{\url{https://mobilitydcat-ap.github.io/mobilitydcatap-ui/}}, which provides the possibility of inserting mobilityDCAT-AP metadata using a form-based interface and obtaining the corresponding RDF representation in different serializations. Users can test and experiment with different metadata to learn how different information should be correctly described according to the mobilityDCAT-AP specification in RDF. The form is compliant with the minimum profile of mobilityDCAT-AP.

\subsection{Hosting and Publication}

The fifth step in the roadmap outlines the necessary actions for hosting and publishing the DCAT-AP extension and the controlled vocabularies created for it. Publication and hosting of mobilityDCAT-AP follow the best practices for Semantic Web resources~\cite{berrueta2008best}.

We selected GitHub for the hosting of mobilityDCAT-AP to facilitate also the collaborative evolution and maintenance of the specification. A GitHub organisation\footnote{\url{https://github.com/mobilityDCAT-AP}} is set up for mobilityDCAT-AP and enables rights management for people and repositories. GitHub pages are leveraged to host artefacts online.

For the publication of the specification, we aimed to decouple the selected namespace from the identified hosting solution for both mobilityDCAT-AP and the associated controlled vocabularies. We evaluated the Data Europe namespaces used by SEMIC metadata specifications, but we opted for the \texttt{w3id} service \footnote{\url{https://github.com/perma-id/w3id.org}} that allows to reserve a specific namespace (while Data Europe assigns a unique identifier to compose the namespace IRI) and to define custom redirects. We selected the namespace \url{https://w3id.org/mobilitydcat-ap} and we implemented redirects for both the specification and the controlled vocabularies. Redirection rules manage a content negotiation mechanism able to serve correct machine-readable (e.g., multiple RDF serializations) or human-readable (HTML for ReSpec) representations of resources. Moreover, redirection enables access to different versions of the same resource that can be accessed composing specific URIs (e.g., version 1.0.0 of mobilityDCAT-AP is accessible at \url{https://w3id.org/mobilitydcat-ap/1.0.0}). 
All the mobilityDCAT-AP resources are published online with a Creative Commons CC-BY license 4.0.

\section{Maintenance and Sustainability}\label{sec:maintenance}

The GitHub repositories of mobilityDCAT-AP will be used for maintenance, leveraging the mechanism of \emph{Issues} to discuss collaboratively. Any user may open an issue on the dedicated repository of a mobilityDCAT-AP resource to provide feedback, propose a change or introduce a new use case to be covered by the specification.
Following best practices from DCAT-AP~\cite{changedcat} we established release cycles for bug fixing, minor and major releases of mobilityDCAT-AP using Semantic Versioning notations. The decision-making process is also defined according to the type of change proposed. It will happen in regular meetings of the maintainers of the specification plus invited users (e.g., proposers of the change to be introduced). Each major version will also go through a public review process in a draft version before being officially released. Following this process, we already released a bug fixing version (1.0.1) and a minor release (1.1.0) that represents the current version of mobilityDCAT-AP. The documentation of the specification keeps track of the changes through a dedicated \emph{Changelog} section.

The maintenance of controlled vocabularies is decoupled from the one of the main specification and each controlled vocabulary may evolve (i.e., changing version) independently from the others and from mobilityDCAT-AP. The current version of each controlled vocabulary is reported in the dedicated repository\footnote{\url{https://github.com/mobilityDCAT-AP/controlled-vocabularies}}. An interesting maintenance aspect of controlled vocabularies is related to the \emph{Update Frequency} vocabulary. We implemented this vocabulary by extending the \emph{Frequency} vocabulary from the EU Publication Office with additional values. Nevertheless, we were then able to contribute our requirements directly to the \emph{Frequency} vocabulary that has been updated to introduce new values. We will evaluate the possibility of contributing to additional controlled vocabularies by the EU Publication Office in the future, to ensure a broader adoption and to guarantee better support of labels for all European languages.

Establishing a NAPCORE organisation that will continue the work done by the NAPCORE project in future years will guarantee long-term support for mobilityDCAT-AP. The European Commission is currently holding an open call to assign funding for the following years.

\section{Adoption}\label{sec:adoption}

The mobilityDCAT-AP adopters repository\footnote{\url{https://github.com/mobilityDCAT-AP/adopters}} provides a comprehensive list of National Access Points (NAPs) and mobility data portals adopting the specification to represent their metadata. The goal of this list is to raise awareness about the specification, track its adoption, and assist potential new adopters by showcasing existing implementations they can use as examples. 
The list is organized by the implementation status and the order in which entries were added. Among the first implementers to deploy the mobilityDCAT-AP specification, we have the NAPs of Denmark (\url{https://du.vd.dk/}), Latvia (\url{https://transportdata.gov.lv/}) and Germany (\url{https://mobilithek.info/}). By visiting the respective data portals, it is indeed possible to download metadata according to the specification. The updated list of NAPs that are currently working on the implementation can be consulted in the repository.

Notably, the EU Delegated Regulation 2017/1926 was amended in 2024 by the EU Delegated Regulation 2024/490~\cite{commission_2024} by specifically stating the importance of metadata for the efficient sharing and reuse of data. In particular, the new Regulation explicitly indicated that Member States should cooperate to agree on metadata requirements, considering the common metadata schema mobilityDCAT-AP and following versions.

In the context of the European Data Spaces~\cite{commission_emds}, the ongoing research project deployEMDS (\url{https://deployemds.eu/}) represents a first step towards the EMDS by (i) deploying operational data spaces and common governance mechanisms in nine pilot sites, and (ii) defining the interlinking layer that will enable the interconnectivity of existing and emerging mobility and transport data spaces. The project is fostering the adoption of mobilityDCAT-AP as a reference metadata profiles for both the data spaces and the interlinking layer.

\section{Challenges and lessons learnt}\label{sec:challenges}

This section reports the challenges and the lessons learnt in the implementation of DCAT-AP and by collecting feedback from early adopters.

\noindent\paragraph{Domain stakeholders involvement is a key aspect} The involvement of domain stakeholders from the initial phases of this project was key to incorporating their requirements and ensuring quick adoption of the metadata specification. Good collaboration between metadata and domain experts is necessary to ensure both perspectives are taken into account in defining the specification.
\noindent\paragraph{Leverage existing resources and support services} The guidelines and tools mentioned in this paper, and followed in the deployment of mobilityDCAT-AP, provide great support in defining a metadata specification and are constantly updated and improved. Moreover, the interaction with the maintainers of other metadata specifications and controlled vocabularies should be leveraged to improve the alignment among parallel specifications. Lastly, a crucial success factor is direct and intense support of early adopters. With a new metadata specification, early adopters have had only little orientation or practice from other deployers. To this end, we offered one-to-one consulting for such adopters, conducted testing of mobilityDCAT-AP export files, provided suggestions for improvement etc.
\noindent\paragraph{Semantic Web technologies require expertise} The advantages of employing a metadata specification based on Semantic Web technologies are well-understood by the relevant stakeholders that struggle with alternative approaches for metadata management. However, generating and managing RDF serialization requires a certain expertise that is often not easy to obtain. We noticed difficulties among the first adopters, not purely in implementing the specification but more generally in generating a valid RDF serialisation of the metadata.  
\noindent\paragraph{Legacy and new implementations} The provided specification offered great help for new implementations of NAPs that did not have to design a metadata schema but could directly refer to mobilityDCAT-AP. Nevertheless, two main aspects still require attention and are related to the interface used to collect metadata (that may influence the quality of the collected information) and the decisions regarding which recommended and optional properties are implemented. Existing NAPs working on the implementation of mobilityDCAT-AP encountered instead more challenges, especially related to the problem of converting already collected metadata and defining mechanisms to request users certain mandatory fields currently not available.
\noindent\paragraph{Maintenance requires a huge effort} Keeping track of issues from users and managing the evolution of the specification requires a huge effort. Well-defined processes should be put in place both to ensure feedback to users and the resolution of the issues with a final decision. The same applies to the generation of a new version of the specification and related vocabularies that should apply consistent workflows to ensure alignment and update of all the related artefacts.


\section{Conclusions}\label{sec:conclusion}

This paper presented mobilityDCAT-AP, a metadata specification to enable the definition of harmonised metadata for data portals dealing with the mobility domain. The main objective of the specification is to enable cross-portal querying of data sources and improve the discoverability of mobility data.
The metadata specification is made available online together with a set of relevant artefacts and resources to support its implementation by relevant stakeholders. Early adopters already deployed the specification in their data portals and a long-term plan for the maintenance of the specification has been identified.
Additionally, we discussed the generic methodology followed for the development of this resource and potentially applicable to define metadata specifications for other domains.
As future work, we plan to evolve mobilityDCAT-AP to accommodate feedback from the community and align with the new DCAT-AP version 3. Additionally, we are planning activities to provide support and training to NAP operators interested in adopting the metadata specification. In this direction, we also evaluating the definition of a \emph{curriculum} that could guide interested stakeholders in gaining the relevant expertise to implement and contribute to mobilityDCAT-AP. Finally, we will focus on guidelines for exposing metadata (e.g., OAI-PMH\footnote{\url{https://www.openarchives.org/pmh/}}) to facilitate harvesting scenarios for metadata exchange across data portals.

\subsubsection{\ackname} This work has been carried out in the context of Sub-working Group 4.4 of the NAPCORE project, an EU co-funded Programme Support Action under Grant Agreement No.\ MOVE/B4/SUB/2020-123/SI2.852232. The Version of Record of this contribution will be published in The Semantic Web - ESWC 2025 proceedings. 


\bibliographystyle{splncs04}
\bibliography{biblio}

\end{document}